\documentclass[amssymb,amsmath,amsfonts,pra]{revtex4-1}
\usepackage{verbatim}
\usepackage{amssymb}
\usepackage{amsmath}
\usepackage{diagbox} 
\usepackage[english]{babel}
\usepackage[latin1]{inputenc}
\usepackage[usenames, dvipsnames]{color}
\usepackage{graphicx}
\usepackage{gensymb}
\usepackage{colortbl}
\usepackage{mathtools}
\usepackage{dsfont}
\usepackage{hyperref}
\usepackage{tikz}
\usepackage{soul}
\DeclarePairedDelimiter\ceil{\lceil}{\rceil}
\DeclarePairedDelimiter\floor{\lfloor}{\rfloor}

\def\ket#1{|#1\rangle}

\def\tr{\mathrm{tr}}
\def\ket#1{\left| #1\right>}

\definecolor{LightCyan}{rgb}{0.8,1,1}
\definecolor{LightRed}{rgb}{1,0.8,0.8}
\definecolor{LightPurple}{rgb}{0.8,0.5,1}

\usepackage{bm}
\usepackage{enumerate}

\newcommand{\braopket}[3]{\left< #1 \vphantom{#2#3} \right| #2 \left| #3 \vphantom{#1#2} \right>}

\begin{document}
\title{Absolutely maximally entangled states, quantum maximum distance separable codes, and quantum repeaters}
\author{Daniel Alsina}
\email{d.alsinaleal@leeds.ac.uk}
\author{Mohsen Razavi}
\affiliation{School of Electronic and Electrical Engineering, University of Leeds, United Kingdom.}

\date{\today}

\begin{abstract}
We address the relation between absolutely maximally entangled (AME) states and quantum maximum distance separable (QMDS) codes by constructing whole families of QMDS codes from their parent AME states. We introduce a reduction-friendly form for the generator set of the stabilizer representation of an AME state, from which the stabilizer form for children codes, all QMDS, can be obtained. We then relate this to optimal codes for one-way quantum repeaters, by minimizing the short-term infrastructure cost as well as the long-term running cost of such quantum repeaters. We establish that AME states provide a framework for a class of QMDS codes that can be used in quantum repeaters.

\end{abstract}

\maketitle

\section{Introduction}

Quantum Internet \cite{Ki08}, viz., the structure via which unknown quantum states can reliably be transferred, heavily relies on the development of some of the most advanced technologies that enable quantum repeaters \cite{Br98,Mu15,Fo10,Mu12}. Early proposals for reliable state transfer rely on teleportation \cite{Br98}, which, itself, requires us to distribute and store high-fidelity entangled states between remote partners in an efficient and scalable way. Most of the experimental efforts on quantum repeaters have so far focused on this approach, albeit a probabilistic version of it. This may prove useful for certain applications, such as quantum key distribution (QKD), which can use post-selection in their procedure. More recent proposals for quantum repeaters go around this limitation by offering an encode-and-go feature in which the original quantum message is encoded into multipartite entangled states, resilient to loss and error, and sent to the next node for error correction and retransmission \cite{Mu15,Fo10,Mu12}. This approach mostly resembles how packets of data are being transmitted across today's communications networks, and can form the ultimate solution to the requirements of a truly working quantum Internet. The required specifications for the codes that enable such a smooth transmission of quantum states is, nonetheless, quite daunting. In this paper, we make an attempt to better understand the underlying structure for efficient codes used in such repeaters, sometimes referred to by third generation quantum repeaters. Along the way we introduce a class of codes that are inherited from an interesting class of states known as absolutely maximally entangled (AME) states \cite{He12,He13a}. We will then explore how such codes can enhance the performance of quantum repeaters that rely on error correction.



Quantum error correction (QEC) is a promising technique to solve the inherent fragility of quantum systems interacting with the environment \cite{Ni00,De13}. In the case of third generation quantum repeaters, there are certain types of errors that can be reduced by the use of quantum error correction codes. One of the key challenges to be addressed by quantum repeaters is the issue of channel loss, which, in the case of flying quantum states carried by photons, can result in {\em erasure} errors. That corresponds to the cases that the photons that are carrying the quantum information will get lost before getting to their destination. Resilience to erasure errors is then one of the key requirements for codes designed to be used in such quantum repeaters. In addition to erasure errors, any quantum circuit that is used for quantum error correction must deal with the operational gate errors. That would possibly require additional protection to offer fault-tolerant operation against both erasure and gate errors. Ideally, both features of correcting possibly a large number of erasure errors and gate errors must be achieved by using a minimum amount of resources. That is, efficiency is another desirable feature of the codes deployed in quantum repeaters.

While several codes have thus far been proposed to be used in third generation quantum repeaters, a systematic approach to selecting a code, based on a desirable set of criteria, is missing. Quantum parity codes were studied first for their simplicity, as they are represented by blocks of several qubits each. Erasure errors can be corrected in such codes as long as each repeater station receives at least a complete block, and at least one qubit for each of the remaining blocks \cite{Mu12}. Quantum parity codes have the property of being Calderbank-Steane-Shor (CSS) codes, and this possibly allows for a simple fault-tolerant implementation. Detailed calculations on an approach based on teleported error correction (TEC) \cite{Kn05} were performed later, showing their ability to correct operational errors as well \cite{Mu14}. But quantum parity codes are not very efficient in terms of the total number of qubits used, and other codes, using fewer resources, were proposed afterwards, e.g., quantum polynomial codes (QPyC) \cite{Gl16,Mu17} and quantum Reed-Solomon codes (QRSC) \cite{Mu18}. An explicit argument on why or when to choose these codes is, however, missing.

\begin{figure}[h!]
\centering
\includegraphics[scale=0.8]{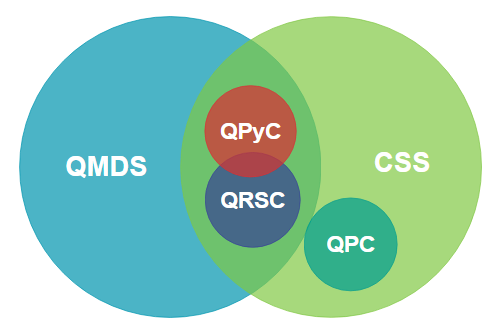}
\caption{Venn diagram showing the relations between the different types of codes mentioned in the paper. QMDS codes are the ones able to correct the maximum number of errors for a fixed amount of resources, and are those related to AME states. CSS codes are specially simple to implement fault-tolerantly. Therefore the intersection of the two is potentially the best region to look for optimal codes for repeaters, like QPyC or QRSC. QPC are instead less efficient.}
\label{venndiagram}
\end{figure}

In this paper, our objective is to find an overarching framework that identifies codes suitable for repeater applications. We are particularly interested in the property of being quantum maximum distance separable (QMDS), as a good indicator of efficiency for our codes. This allows for maximum possible number of error correction given fixed resources. As in previous proposals, we also associate simplicity to the CSS family of codes. Figure~\ref{venndiagram} shows a Venn diagram for the relationship between CSS and QMDS codes, and how each of the above example codes would fit into the diagram. QPyC and QRSC, in particular, seem to offer both efficiency and relative simplicity as they are both QMDS and CSS codes. While CSS codes are well studied in the literature, here we turn our focus to  a group of QMDS codes that are related to AME states. 

AME states belong to a class of multipartite entangled states that are characterized by having a completely mixed state in all their balanced bipartitions, hence offering maximum possible average entropy. An interesting point that connects AMEs to error correction is their usage in quantum secret sharing (QSS). In threshold QSS, a dealer distributes a secret quantum state among $N$ parties such that a subset $K$ of them have to cooperate to reconstruct the secret, whereas $K-1$ parties alone do not get any information about it. But this also means that $K$ of them are able to recover the secret regardless of what the other parties do, including destroying their shares, which is an equivalent situation to a code that can correct for a certain number of erasure errors. AME states are shown to be optimal for threshold QSS \cite{He12}, as any balanced bipartition of them is a completely mixed state, that is, no information can be gained about the state if less than $N/2$ parties join forces. This suggests a relationship between AME states and erasure errors \cite{He13a,Go15a,Ra18}, which could be used among other things in quantum repeaters.  


AMEs are linked to QMDS codes as well. In particular, AMEs are shown to be equivalent to a special QMDS code in Ref. \cite{Sc04}, which has only one codeword. Being the only codeword in the code space may not be attractive for communication purposes as we need to encode different qubits into different codewords. However one can obtain a family of useful QMDS codes from an AME, as shown in Refs. \cite{Ca98,Ra98,Ke06}. These families have been studied thoroughly in a recent paper \cite{Hu19}, where they are used to find weight enumerators and bounds of QMDS codes.

QMDS codes that are considered in this work are part of the class of additive codes \cite{Ca98}, which can be described using the stabilizer formalism \cite{Go97}. Interestingly, all hitherto known AMEs can also be described with stabilizers. In this work we propose a new convenient way to write AMEs in this language. This form makes it easy to obtain the families of QMDS codes from their ``parent'' AME state.  We prove with our new construction that the ``children'' codes do meet the required conditions for being a QMDS code, in all qudit dimensions that are a prime integer or power of a prime integer. Note that the existence of such children codes had previously been demonstrated only in the qubit space \cite{Go97}. We then show how QMDS codes can be relevant to quantum repeaters, and make some comparisons between the performance of different types of codes.

The rest of this paper is organized as follows. In Sec.~\ref{definitions}, we provide our terminology for the commonly used tools in this paper, including QEC codes, QMDS codes, AME states and the stabilizer formalism. Section~\ref{observation} introduces the idea of obtaining a children code from an AME state with an explicit example. Section~\ref{construction} is the core of the work: we write AME stabilizers in a convenient way such that the stabilizer representation for children codes can be found in a straightforward way. A proof of their fulfillment of the requirements of a code is given, in two cases: for qudit dimensions that are prime, and for dimensions that are power of a prime. Section~\ref{repeaters} explores how  QMDS codes are promising candidates for the implementation of one-way quantum repeaters and shows some comparative results for key generation rates of a QKD setup and cost coefficients of different types of codes. We conclude the paper in Sec.~\ref{Sec:Diss}.

\section{Terminology and Notations}
\label{definitions}
Here, we summarize the notation used in this paper.

{\it Quantum error-correcting code (QECC)}:  A subspace $\mathcal{C}$ spanned by a set ${\ket {\psi_m}}, m \in \{1,...,K\}$, of orthonormal states is a $((n,K,d))_q$ QECC, i.e. a code that encodes a logical space of dimension $K$ into $n$ physical qudits of dimension $q$, if it obeys the Knill-Laflamme conditions \cite{Kn97}
\begin{equation}
\forall m,m' \in \{1,...,K\}: \braopket {\psi_m} {E^{\dagger}F} {\psi_{m'}}=f(E^{\dagger}F)
\delta_{m,m'},
\label{kl}
\end{equation}
for all error operators $E$ and $F$ with $wt(E^\dagger F) < d$, where $wt$ is the $weight$ of an operator, and the parameter $d$ is the $distance$ of the code; see Ref. \cite{Kn97} for full definition of these notions. We will define such features of a code for our cases of interest in the stabilizer formalism described later in this section. In Eq. \eqref{kl}, $f$ is a function acting on the error operators and $\delta_{m,m'}$ is the Kronecker delta function \cite{Kn97}.

In this paper we will focus on {\it additive} codes \cite{Ca98}, a subclass of codes characterized by having a certain group structure that is best captured by the stabilizer formalism. They incorporate the restriction that $K=q^k$, where now $k$ can be interpreted as the length of the unencoded message in qudits of dimension $q$. Such codes are represented with the notation $[\![n,k,d]\!]_q$. \\

{\it Quantum maximum-distance separable (QMDS) code}: A QECC is a QMDS code if its parameters fulfill the {quantum Singleton bound (QSB)}, i.e., $n-k \ge 2(d-1)$, with equality. They are optimal in the sense that they achieve the maximum possible distance for a given $n-k$. 
Note that when $n$ and $k$ have different parity, it will not be possible to saturate the Singleton bound. We will call the codes with the maximum possible distance allowed by the bound in that situation {\it suboptimal QMDS}.

QMDS codes are not guaranteed to exist for every possible combination of $n$, $k$ and $q$. A list of maximum possible distances for all binary codes of $n$ up to 100 appears in Ref. \cite{Gr15}. \\

{\it Absolutely maximally entangled (AME) state}: An AME($n$,$q$) state
$|\psi\rangle \in {\cal H}$, for ${\cal H}=(\mathds{C}^q)^{\otimes n}$, of $n$ qudits of local dimension $q$, is a state such that all its balanced partitions ${\cal A}=(\mathds{C}^q)^{\otimes \floor{\frac{n}{2}}}$, ${\cal H}={\cal A} \otimes \bar{\cal A}$,
carry maximal entropy, i.e.,
\begin{equation}
  S(\rho_{\cal A})=\floor{\frac{n}{2}} \log_2 q, \qquad \forall {\cal A}  \ ,
\end{equation}
with $\rho_{\cal A} = \tr_{\bar{\cal A}}{(|\psi\rangle \langle \psi|)}$. This is tantamount to requiring that the reduced density matrix to $r$ qudits, $\rho_r$, to be proportional to the identity, with $r$ ranging from $1$ to $\floor{\frac{n}{2}}$. That is,
\begin{equation}
\rho_r=\frac{1}{q^r} \mathbb{I}_{q^r}, \qquad \forall r\leq \floor{\frac{n}{2}}\ .
\end{equation}

AMEs do not exist in all possible Hilbert spaces. For qubits they only exist for 2, 3, 5 and 6 parties, whereas it has been shown that AMEs do not exist for 4 \cite{Hi00}, 7 \cite{Hu17} and more than 8 qubits \cite{Sc04}. A thorough table on the existence of AMEs for many different Hilbert spaces can be found in Ref. \cite{Hu18a} and in Fig. 2 in Ref. \cite{Hu18b}. In Hilbert spaces where AME states do not exist, closest states to an AME state may be defined.

A convenient way to characterize QMDS codes and AMEs is by using the stabilizer formalism \cite{Go97}, which we define now. It is suspected that AME states can always be described in terms of the stabilizer formalism. This has not been proven, but it is true for all known examples \cite{Hu18c}. In this work we only consider AME states that have stabilizer representations. \\

{\it Stabilizer formalism}: An operator $S$ is said to stabilize a state $\ket{\phi}$ if
\begin{equation}
S\ket{\phi} = \ket{\phi}.
\end{equation}
In quantum error correction, stabilizer operators are often described as tensor products of Pauli operators, or the generalization of them to $q$ dimensions, which for the moment we assume is a prime number. The generalized Pauli operators, $X_a$ and $Z_a$, for prime $q$ and $a \in \mathbb{Z}_q$, act on the computational basis states as follows
\begin{eqnarray}
&&X_a \ket{j} = \ket{j+a \hspace{2mm} \rm{mod} \hspace{2mm} \it{q}} \nonumber \\ 
&&Z_a \ket{j} = \omega^{aj}\ket{j},
\label{pauliop}
\end{eqnarray}
with $\omega=e^{2\pi i/q}$. Note that for prime $q$, $X \equiv X_1$ and $Z \equiv Z_1$ are unitary and traceless, and we have $X_a = X^a$ and $Z_a = Z^a$.  Furthermore, $X^q = Z^q = I$, $ZX=\omega XZ$, and $XZ=\omega^{q-1}ZX$. Any stabilizer operator written as a tensor product of $n$ elements from this basis defines a subspace of the Hilbert space ${\cal H}=(\mathds{C}^q)^{\otimes n}$ with the states that have eigenvalue +1 for that operator. Any state in ${\cal H}$ can be stabilized by a total of $q^n$ different stabilizer operators of this form. Such stabilizers can be generated from a set of $n$ independent operators, known as the generator set, as products of powers of its elements \cite{Go97}.

An additive QECC is a subspace of states, and an $[\![n,k,d]\!]_q$ code can be described by $n-k$ generators, for $q$ prime. The generators of the stabilizer space of that code must fulfill the following conditions:
\begin{enumerate}
\item Independence: $g_i \neq \prod_{j=1,j\neq i}^{n-k} g_j^{m_j}
\hspace{10mm} \forall m_j=0,1,...,q-1$ and $ i=1,...,n-k$.
\item Commutation:
$[g_i,g_j]=0 \hspace{10mm} \forall i,j=1,...,n-k$
\item Distance:
$\forall E$ with $1\leq wt(E) \leq d-1$, $\exists i$ such that $[E,g_i]\neq 0$,
\end{enumerate}
where operators $E$ are also written in terms of Pauli operators, and they represent the possible errors affecting an encoded state. The weight of an error, $wt(E)$, is the number of sites of the error operator with elements different from the identity. The distance condition allows us to detect errors with weight smaller than $d$, the distance of the code, by identifying generators that do not commute with that error. \\

An interesting property of AME states is that an AME($n$,$q$) is equivalent to a $[\![n,0,\floor{\frac{n}{2}}+1]\!]_q$ code \cite{Sc04}. This is a code with only one codeword, hence it cannot be directly used for reliable communication. In the next section we give an example of how an AME state can be used as a parent code for a family of QMDS codes. We will then extend this technique to all AME states with valid stabilizer representations in Sec. \ref{construction}.

\section{Observation}
\label{observation} 

Let us consider the AME(4,3) state given by \cite{Go15a} 
\begin{eqnarray}\label{eq:ame43}
\ket{\psi}=&&
\frac{1}{3}\sum_{i,j,=0,1,2} |i\rangle|j\rangle |i+j\rangle |i+2 j\rangle \nonumber\\  
= &&\frac{1}{3}\left( \ket{0000}+\ket{0112}+\ket{0221}+\ket{1011}+\ket{1120}+\ket{1202}+\ket{2022}+\ket{2101}+\ket{2210}\right) \nonumber\\ 
= &&\frac{1}{3}\left( \ket{0} \left( \ket{000}+\ket{112}+\ket{221} \right) +\ket{1} \left( \ket{011}+\ket{120}+\ket{202} \right) +\ket{2} \left( \ket{022}+\ket{101}+\ket{210}\right) \right)
\ ,
\end{eqnarray}
where the sums $i+j$ and $i+2j$ are taken modulo $q=3$, and the state is written in the standard computational basis. By proposition 3 in Ref. \cite{Sc04}, $|\psi\rangle$ is equivalent to a $[\![4,0,3]\!]_3$ code. We now try to build a $[\![3,1,2]\!]_3$ code by treating the first qutrit in $|\psi\rangle$ as the message part, and encoding it with the projection of $|\psi\rangle$ onto the message component. That is, from Eq. \eqref{eq:ame43}, the logical codeword for qutrit $i$ is given by  
\begin{equation}\label{eq:qecc312}
\ket{i_L} =
\frac{1}{\sqrt{3}}\sum_{j,=0,1,2} |j\rangle |i+j\rangle |i+2 j\rangle \hspace{10mm} {\rm for} \hspace{2mm} i=0,1,2 ,
\end{equation}
or, equivalently,
\begin{eqnarray}\label{eq:qecc312exp}
&&\ket{0_L} =\frac{1}{\sqrt{3}}\left( \ket{000}+\ket{112}+\ket{221}\right) \nonumber \\ 
&&\ket{1_L} =\frac{1}{\sqrt{3}}\left( \ket{011}+\ket{120}+\ket{202}\right) \nonumber \\ 
&&\ket{2_L} =\frac{1}{\sqrt{3}}\left( \ket{022}+\ket{101}+\ket{210}\right)
\ .
\end{eqnarray}
We claim that the codewords in \eqref{eq:qecc312exp} indeed form a $[\![3,1,2]\!]_3$ code, which is QMDS. This is an example of a general construction of reduced codes, whose existence has been proven in Theorem 6, part b, in Ref. \cite{Ca98}, Theorem 20 in Ref. \cite{Ra98}, Lemma 70 in Ref. \cite{Ke06} or Theorem 2 in Ref. \cite{Hu19}. In next section, we give an analytical technique that can be applied to prove this and a much broader set of claims within the stabilizer formalism. 

\section{Reduction-friendly stabilizer form for AME states}
\label{construction}

In this section we show how to start from the stabilizer representation of an AME state and obtain the stabilizer form of all possible children codes, systematizing the observation made in the previous section. All these children codes turn out to be QMDS as well. This will provide us with a recipe to generate efficient QMDS codes that inherit strong erasure error correction features. Part of our construction and proof is inspired by Section 3.5 of Gottesman's thesis \cite{Go97}. In there, the author explains how to obtain an $[\![n-1,k+1,d-1]\!]_2$ qubit code from any $[\![n,k,d]\!]_2$ qubit code, giving in detail the necessary manipulations. In our construction, we extend this technique to non-qubit codes of prime, or power of a prime, dimensions. We provide a concrete reduction-friendly form for an AME from which it is easy to automatically obtain the whole family of children codes. The proposed techniques can possibly be generalized to find the stabilizer form for a $[\![n-k',k+k',d-k']\!]_q$ code, with $k'\leq d-1$, from that of an original $[\![n,k,d]\!]_q$ code, even if the original code is not QMDS. 

\subsection{Construction for prime $q$}
\label{coreconstruction}

For an AME($n$,$q$) state, with $q$ a prime number, there must exist $n$ generators for the equivalent $[\![n,0,\floor{n/2}+1]\!]_q$ code, whenever a valid stabilizer representation exists. We will show that the stabilizer generator set for these codes, when $n$ is even, can always be written as in Fig.~\ref{canst}(a), and, when $n$ is odd, as in Fig.~\ref{canst}(b), where an extra generator, containing identities in all the left $\floor{n/2}$ sites, has been added. In Fig.~\ref{canst}, each generator is represented by the tensor product of $n$ Pauli operators, whose first $\floor{n/2}$ of them are specified in the given form. The other $\ceil{n/2}$ Pauli operators in each generator would be specific to the particular AME state of interest. The structure in Fig.~\ref{canst} is of a convenient form because it easily allows us to obtain whole families of QMDS codes from their parent AME state, as we show below in Proposition 1. We will refer to the forms in Fig.~\ref{canst} as the {\em reduction-friendly} stabilizer forms of an AME state. We will show in Proposition 2 that such a reduction-friendly form exists for any AME state with a valid stabilizer representation. Throughout this subsection, we assume that $q$ is a prime number. 

\begin{figure}[h!]
\begin{tabular}{c | c c c c c | c}
(a) & \multicolumn{5}{c}{$n/2$ sites} & $n/2$ sites \\
\hline
$g_1$ & $I$ & $I$ & $...$ & $I$ & $Z$ & $.................$\\
$g_2$ & $I$ & $I$ & $...$ & $I$ & $X$ & $.................$\\
$g_3$ & $I$ & $I$ & $...$ & $Z$ & $I$ & $.................$\\
$g_4$ & $I$ & $I$ & $...$ & $X$ & $I$ & $.................$\\
$\vdots$ & $\vdots$ & $\vdots$ & $\vdots\vdots\vdots$ & $\vdots$ & $\vdots$ & $\vdots\vdots\vdots\vdots\vdots\vdots\vdots\vdots\vdots\vdots\vdots\vdots\vdots\vdots\vdots\vdots\vdots$\\
$g_{n-3}$ & $I$ & $Z$ & $...$ & $I$ & $I$ & $.................$\\
$g_{n-2}$ & $I$ & $X$ & $...$ & $I$ & $I$ & $.................$\\
$g_{n-1}$ & $Z$ & $I$ & $...$ & $I$ & $I$ & $.................$\\
$g_{n}$ & $X$ & $I$ & $...$ & $I$ & $I$ & $.................$
\\
\hline
\end{tabular}
\quad
\begin{tabular}{c | c c c c c | c}
(b) & \multicolumn{5}{c}{$\floor{n/2}$ sites} & $\ceil{n/2}$ sites \\
\hline
$g_1$ & $I$ & $I$ & $...$ & $I$ & $I$ & $.................$\\
$g_2$ & $I$ & $I$ & $...$ & $I$ & $Z$ & $.................$\\
$g_3$ & $I$ & $I$ & $...$ & $I$ & $X$ & $.................$\\
$g_4$ & $I$ & $I$ & $...$ & $Z$ & $I$ & $.................$\\
$g_5$ & $I$ & $I$ & $...$ & $X$ & $I$ & $.................$\\
$\vdots$ & $\vdots$ & $\vdots$ & $\vdots\vdots\vdots$ & $\vdots$ & $\vdots$ & $\vdots\vdots\vdots\vdots\vdots\vdots\vdots\vdots\vdots\vdots\vdots\vdots\vdots\vdots\vdots\vdots\vdots$\\
$g_{n-3}$ & $I$ & $Z$ & $...$ & $I$ & $I$ & $.................$\\
$g_{n-2}$ & $I$ & $X$ & $...$ & $I$ & $I$ & $.................$\\
$g_{n-1}$ & $Z$ & $I$ & $...$ & $I$ & $I$ & $.................$\\
$g_{n}$ & $X$ & $I$ & $...$ & $I$ & $I$ & $.................$\\
\hline
\end{tabular}
\caption{Reduction-friendly generator forms of AMEs with stabilizer representations. Each row represents the tensor product of $n$ Pauli operators that form a generator. For instance, in (a), $g_1 = I \otimes I \otimes ... \otimes I \otimes Z \otimes \bar g_1$, where $\bar g_1$, represented by dots, is the tensor product of the other $n/2$ Pauli operators. (a) Reduction-friendly generator set for an $AME(n,q)$ for $n$ even. The left part, with $n/2$ components, guarantees independence of the generators, whereas the remaining $n/2$ Pauli operators must be found by imposing the remaining conditions of commutation and distance for a generator set. (b) When $n$ is odd, an extra generator must be added, containing identities in all the left $\floor{n/2}$ sites, as in $g_1$. }
\label{canst}
\end{figure}


{\it Proposition 1}. For an even integer $n$, if an $[\![n,0,n/2+1]\!]_q$ code has stabilizer generators in the form given by Fig.~\ref{canst}(a), then we can obtain the generator set for its QMDS child code, $[\![n-1,1,n/2]\!]_q$, by removing the last two generators and the first column in Fig.~\ref{canst}(a) of the $[\![n,0,n/2+1]\!]_q$ code. That is, generators $g_1, \ldots, g_{n-2}$, once stripped of operator $I$ in their first site, will form the generator set for $[\![n-1,1,n/2]\!]_q$.
\\

{\it Proof}. Here, we extend the proof presented in Sec. 3.5 of Ref. \cite{Go97}, for the special case of qubits, to codes of a prime dimension $q$. Let us refer to the suggested generators for $[\![n-1,1,n/2]\!]_q$ by $g^*_1,\ldots,g^*_{n-2}$. In order to show that these operators will form the generator set for $[\![n-1,1,n/2]\!]_q$, we have to show that they satisfy the independence, commutation, and distance criteria. The first two follow from the fact that $g_1,\ldots,g_{n-2}$ are independent and mutually commute. Let us focus on the distance criterion then. Suppose the code generated by $g^*_1,\ldots,g^*_{n-2}$ has distance $p$. That is, $p$ is the smallest weight for which there is an error $E$ on sites 2 to $n$ that commutes with all generators $g^*_1,...,g^*_{n-2}$. By quantum Singleton bound, we know that $p \leq n/2$. We now show that $p \geq n/2$ as well, which proves that $p=n/2$. That is, the code generated by $g^*_1,...,g^*_{n-2}$ is a QMDS code.

In the following, we search for an error operator $E''$ that commutes with $g_1,\ldots,g_{n}$. Consider the error operator $E'=I \otimes E$, {where we remind that $E$ is an error operator of weight $p$ that commutes with $g^*_1,...,g^*_{n-2}$}. $E'$ will commute with generators $g_1,\ldots,g_{n-2}$ as they have only identities on their first site. If it also commutes with $g_{n-1}$ and $g_{n}$, then we choose $E''=E'$. This implies that the distance for the original code, $n/2+1$, must be less than or equal to $p$. This contradicts the Singleton bound for the truncated code, and cannot be the case. This implies that one or both of $g_{n-1}$ and $g_{n}$ would not commute with $E'$. In fact, given that $E'$, $g_{n-1}$, and $g_{n}$ are all tensor products of Pauli operators, we can find coefficients $m_{n-1}$ and $m_n$ such that 
\begin{eqnarray} 
g_{n-1}E' && = \omega^{m_{n-1}} E'g_{n-1}, \\ \nonumber
g_{n}E' && = \omega^{m_{n}} E'g_{n},
\end{eqnarray}
where $m_{n-1}, m_n \in \{0,...,q-1\}$, and not both equal to zero. We can then define the error operator $E'' = X^{q-m_{n-1}} Z^{q-m_{n}} \otimes E$, which commutes will all $g_1,\ldots,g_{n}$. This implies that the distance of the original code, $n/2+1$, must be less than or equal to the weight of $E''$, i.e., $p+1$. That is, $p \geq n/2$, which completes our proof.  $\square$
\\

{\it Corollary 1.1}. The above procedure can be iterated up to $n/2-1$ times to produce a whole family of $[\![n-k,k,n/2+1-k]\!]_q$ codes, with $k={0,...,n/2-1}$.
\\

{\it Corollary 1.2}. The same reasoning can be applied to the AME state of Fig.~\ref{canst}(b) of $n$ odd, to produce a whole family of $[\![n-k,k,\floor{n/2}+1-k]\!]_q$ codes, with $k={0,...,\floor{n/2}-1}$.
\\

{\it Corollary 1.3}. Note that in the proof of Proposition 1, we did not make any use of the particular format of $g^*_1,...,g^*_{n-2}$ as given by Fig.~\ref{canst}(a). We only used the particular form of the first column in Fig.~\ref{canst}(a). The claim in Proposition 1 still holds so long as the first column has the desired form. This observation is important as we later use it in the proof of Proposition 2. 
\\

We are now going to prove that it is always possible to bring an AME state with a stabilizer representation to the reduction-friendly forms in Fig.~\ref{canst}, and, in such cases, it is always possible to apply Proposition 1 to obtain a whole family of $[\![n-k,k,\floor{n/2}+1-k]\!]_q$ codes for $k=\{1,...,\floor{n/2}\}$. 
\\

{\it Proposition 2}. For a stabilizer $AME(n,q)$ state, or its equivalent $[\![n,0,n/2+1]\!]_q$ code, with $n \geq 2$ even and $q$ prime, we can find stabilizer generators such that, for $j = \{1,...,n/2\}$, 

 \[ \begin{cases} 
      g_{i,j} = Z & {\rm if} \hspace{2mm} i=n+1-2j \\
      g_{i,j} = X & {\rm if} \hspace{2mm} i=n+2-2j \\
      g_{i,j} = I & {\rm elsewhere} 
   \end{cases}
\]
where $g_{i,j}$ refers to the Pauli operator on site $j$ of generator $g_i$, as shown in Fig.~\ref{canst}(a). 
\\

{\it Proof}. {\color{blue} We will proceed in a similar fashion to Gaussian elimination over $Z_q \times Z_q$}. If $AME(n,q)$ is a stabilizer state, then it should have $n$ generators. Let us denote these generators by $g_1, \ldots, g_n$. 

{\it Claim 1:} There are at least two generators in the set whose first Pauli element is not identity. Let us see why: if for all $g_1$ to $g_n$, the first site is the identity operator, then the error operator $X \otimes I^ {\otimes n-1}$ commutes with all generators. This implies that the distance of the code is 1, whereas $n/2+1 \geq 2$. Suppose $g_n$ is an operator with non-identity operator in its first site. If $g_n$ is the only generator with this property, then the error operator $g_{n,1} \otimes I^ {\otimes n-1}$ would commute with all generators, and we again reach the same distance contradiction. It follows that there are at least two generators with non-identity operators in their first site. For the rest of the proof, we assume that we have relabelled the generators so that $g_{n-1}$ and $g_n$ are two such operators.

Now, consider the most general case for the first column of $g_{n-1}$ and $g_n$:
\begin{eqnarray} 
g_{n-1,1} && = X^a Z^b \\ \nonumber
g_{n,1} && = X^c Z^d
\end{eqnarray}
where $a,b,c,d \in \mathbb{Z}_q$. We take the following four steps to reach the form given in Fig.~\ref{canst}(a).

{\it Step 1: Elimination of $Z^d$ to obtain ${g'}_{n,1} = X^{c'}$}. As we are free to substitute a generator by a power of itself multiplied by powers of the others, let us make the following transformation
\begin{equation}
g'_n = (g_{n-1})^\beta (g_{n})^\delta,
\label{powers}
\end{equation}
where $\beta, \delta \in \mathbb{Z}_q$. To achieve $g'_{n,1} = X^{c'}$, up to a constant factor in power of $\omega$, we need to find a solution to the following equation
\begin{equation}
\beta b + \delta d = 0\hspace{1mm} ({\rm mod}\hspace{1mm}q),
\label{bezout}
\end{equation}
which is equivalent to finding an integer $m$ such that
\begin{equation}
\beta b + \delta d = mq .
\label{bezoutkq}
\end{equation}
According to B\'ezout identity \cite{Be1779}, $\beta$ and $\delta$ will always exist provided that $mq$ is a multiple of the greatest common divisor of $b$ and $d$. But this is easy to fulfill, as $m$ can be chosen such that the equation has a solution.

If this first step produces $c'=0$, and, therefore, $g'_{n,1} = I$, this means that $g_{n-1}$ and $g_{n}$ are not independent on the first site. We then have to permute generators and select another $g_n$ that is independent with $g_{n-1}$ on the first site. It is guaranteed to exist, otherwise the stabilizers would not define a code of distance larger than 1, as we showed under Claim 1. \\

{\it Step 2: Transformation of $X^{c'}$ to $X$}. That means we need to find $\gamma$ such that $c'\gamma = 1$. Therefore an inverse for $c' \neq 0$ is needed. Given that $q$ is prime, $\mathbb{Z}_q$ is a field, meaning that every element in it has a multiplicative inverse. \\

{\it Step 3. Transformation of the rest of the first column.} The same procedure can be carried to transform $g_{n-1,1}=X^a Z^b$ to $g'_{n-1,1} = Z^{b'}$ and then to $g''_{n-1,1} = Z$, all up to a constant factor, which can be ignored. Then, we can easily make all the other generators to have identities on the first site by taking products of themselves with the last two generators, now using the fact that every element in $\mathbb{Z}_q$ has an additive inverse, which is true for any $q$. \\

{\it Step 4. Iteration over the rest of the sites.} Now that the generator table has the same form to Fig.~\ref{canst}(a) in its first column, we can apply Corollary 1.3 to remove the last two generators ($g_{n-1}$ and $g_n$) and the whole first column, and obtain an $[\![n-1,1,n/2]\!]_q$ code. We can then repeat Steps 1--3 for the truncated code, generating $g^*_{n-3,1} = Z$ and $g^*_{n-2,1} = X$, and all other $g^*_{i,1} = I$. We can then go back to the original code and obtain $g_{n-1,2} = I$ and $g_{n,2} = I$ using powers of $g_{n-3}$ and $g_{n-2}$, without worrying about changing the first column as those generators have identity operators in their first site. This will bring the first two columns and the last four rows of the code in the reduction-friendly form of Fig.~\ref{canst}(a). One can then iterate this procedure until column $n/2$, transforming the whole code to its reduction-friendly form. $\square$ 
\\

{\it Corollary 2.1}. The case of $n$ odd, which represents a $[\![n,0,\floor{n/2}+1]\!]_q$ code, can be treated equivalently to the $[\![n-1,0,(n-1)/2+1]\!]_q$ code of an even number of parties. Applying Proposition 2 to the $[\![n-1,0,(n-1)/2+1]\!]_q$ code we can then obtain the corresponding form for generators $g_2,\dots,g_n$ in Fig.~\ref{canst}(b). We need however an extra generator $g_1$, which can be set to have identities in all sites of the first half, as in Fig.~\ref{canst}(b), as we do not need it to have any special form to guarantee the distance criterion.
\\



Figure~\ref{2022} gives all families of QMDS codes obtained from a parent AME state of $q=2$. We have applied our technique to find the reduction-friendly form for $AME(2,2)$, $AME(3,2)$, $AME(5,2)$ and $AME(6,2)$. In the latter two cases, we have then obtained the generator set for the corresponding children codes as well.

\begin{figure}
\begin{tabular}{c | c | c }
\multicolumn{3}{c}{$[\![2,0,2]\!]_2$}\\
\hline
$g_1$ & $Z$ & $Z$\\
$g_2$ & $X$ & $X$\\
\hline
\end{tabular}
\quad
\begin{tabular}{c}
,
\end{tabular}
\quad
\begin{tabular}{c | c | c c }
\multicolumn{4}{c}{$[\![3,0,2]\!]_2$}\\
\hline
$g_1$ & $I$ & $Z$ & $Z$\\
$g_2$ & $Z$ & $I$ & $Z$\\
$g_3$ & $X$ & $X$ & $X$\\
\hline
\end{tabular}
\quad
\begin{tabular}{c}
,
\end{tabular}

\begin{tabular}{c | c c | c c c}
\multicolumn{6}{c}{$[\![5,0,3]\!]_2$}\\

\hline
$g_1$ & $I$ & $I$ & $X$ & $X$ & $X$\\
$g_2$ & $I$ & $Z$ & $Z$ & $I$ & $Z$\\
$g_3$ & $I$ & $X$ & $Z$ & $I$ & $XZ$\\
$g_4$ & $Z$ & $I$ & $Z$ & $Z$ & $I$\\
$g_5$ & $X$ & $I$ & $Z$ & $XZ$ & $I$\\
\hline
\end{tabular}
\quad
\begin{tabular}{c}
$\rightarrow$
\end{tabular}
\quad
\begin{tabular}{c | c | c c c}
\multicolumn{5}{c}{$[\![4,1,2]\!]_2$}\\
\hline
$g_1$ & $I$ & $X$ & $X$ & $X$\\
$g_2$ & $Z$ & $Z$ & $I$ & $Z$\\
$g_3$ & $X$ & $Z$ & $I$ & $XZ$\\
\hline
\end{tabular}

\begin{tabular}{c | c c c | c c c}
\multicolumn{7}{c}{$[\![6,0,4]\!]_2$}\\
\hline
$g_1$ & $I$ & $I$ & $Z$ & $X$ & $Z$ & $Z$\\
$g_2$ & $I$ & $I$ & $X$ & $Z$ & $XZ$ & $XZ$\\
$g_3$ & $I$ & $Z$ & $I$ & $Z$ & $X$ & $Z$\\
$g_4$ & $I$ & $X$ & $I$ & $XZ$ & $Z$ & $XZ$\\
$g_5$ & $Z$ & $I$ & $I$ & $Z$ & $Z$ & $X$\\
$g_6$ & $X$ & $I$ & $I$ & $XZ$ & $XZ$ & $Z$\\
\hline
\end{tabular}
\quad
\begin{tabular}{c}
$\rightarrow$
\end{tabular}
\quad
\begin{tabular}{c | c c | c c c}
\multicolumn{6}{c}{$[\![5,1,3]\!]_2$}\\
\hline
$g_1$ & $I$ & $Z$ & $X$ & $Z$ & $Z$\\
$g_2$ & $I$ & $X$ & $Z$ & $XZ$ & $XZ$\\
$g_3$ & $Z$ & $I$ & $Z$ & $X$ & $Z$\\
$g_4$ & $X$ & $I$ & $XZ$ & $Z$ & $XZ$\\

\hline
\end{tabular}
\quad
\begin{tabular}{c}
$\rightarrow$
\end{tabular}
\quad
\begin{tabular}{c | c | c c c}
\multicolumn{5}{c}{$[\![4,2,2]\!]_2$}\\
\hline
$g_1$ & $Z$ & $X$ & $Z$ & $Z$\\
$g_2$ & $X$ & $Z$ & $XZ$ & $XZ$\\

\hline
\end{tabular}
\caption{Examples of reduction-friendly forms for AME states in the qubit space: $AME(2,2)$, $AME(3,2)$, $AME(5,2)$, and its child code $[\![4,1,2]\!]_2$, as well as $AME(6,2)$, and its two QMDS children codes.}
\label{2022}
\end{figure}


It must be noted that this technique of suppressing generators and columns from our reduction-friendly generator set of Fig.~\ref{canst} is mathematically equivalent to the procedure for finding the partial trace of a stabilizer state in what is called the row-reduced echelon form in Ref.~\cite{Au05}. This shows that the children codes from an AME state can also be interpreted as {the support of} their mixed stabilizer states obtained from successive partial traces of the original state.

Also note that other QMDS codes of $k\neq0$ might exist even if their parent code with $k=0$ does not. In the table of Ref.~\cite{Gr15}, it can be checked that there are some binary QMDS codes, e.g., $[\![6,4,2]\!]_2$ code, that do not have an AME parent. Another example is for qutrits: whereas the AME of 12 qutrits does not exist, the $[\![8,4,3]\!]_3$ code, as shown in Theorem 14 of Ref. \cite{Gr04}, does exist. It is however worth mentioning that a very recent paper shows that the existence of a $[\![n-1,1,n/2]\!]_q$ code implies the existence of the $[\![n,0,n/2+1]\!]_q$ for $n$ even \cite{Hu19}.

An important discussion that has been left out so far is how to obtain the original stabilizer expression for the AME state, or another parent code. Such stabilizer representations are known for certain cases, but there is no known systematic and/or efficient way {to find all of them}. An attractive idea is to take our reduction-friendly form of Fig.~\ref{canst} for a desired $AME(n,q)$ state and find the right-hand side operators by imposing the commutation and distance criteria for the code, as the first condition, i.e., independence, is guaranteed by the left-hand side. Although this is better than searching for AME states blindly, without a smart strategy to find the right hand side of the code, the computation time will still increase very quickly with $n$ and $q$. 
Another possibility, not necessarily efficient but which provides valuable insight, goes along the way of using the equivalence between graph states and stabilizer states \cite{Sc02,Gr02}. In the end, it amounts to check that a certain number of determinants, with the number of operations growing exponentially with $n$ and $q$, are different from zero. Reference~\cite{He13b} provides a systematic method to find new AME states using graphs.

\subsection{Extension to power-prime $q$}
In this section, we extend the results obtained in Sec.~\ref{coreconstruction} to the case of $q=p^m$ for a prime number $p$ and a positive integer $m$. 
There are several reasons that the previous treatment may not work for non-prime $q$. At the core of them is the mathematical fact that ring $\mathbb{Z}_q$ is not a field when $q$ is not a prime number. In particular, this implies that not all elements in $\mathbb{Z}_q$ have a multiplicative inverse, and that, for example, invalidates Step 2 in the proof of Proposition 2. Different ways have been proposed to circumvent the difficulties of the stabilizer formalism for non-prime $q$. Here, we adopt the approach in Refs.~\cite{Ke06,Go14,Go15b, Gr03}, which give a solution for the special case of $q=p^m$, where Galois field $GF(p^m)$ can be used instead of $\mathbb{Z}_q$. We will not deal here with the more general case of $q \neq p^m$.

The first step to define stabilizer operators for a power-prime $q$ is to extend the definitions in Eq.~\eqref{pauliop} for generalized Pauli operators. Suppose $\alpha$ is a {\it primitive} element of $GF(p^m)$ \cite{Li97}, such that $\alpha^{p^m-1}=1$. We can then write $GF(p^m) = \{0,1,\alpha,\alpha^2,...,\alpha^{p^m-2}\}$. In such a setting, and for $j, a \in GF(p^m)$, the generalized Pauli operators $X_a$ and $Z_a$ satisfy the following relationships 
%
\begin{eqnarray}
&&X_a \ket{j} = \ket{j+a \hspace{2mm} \rm{mod} \hspace{2mm} \it{q}}, \nonumber \\ 
&&Z_a \ket{j} = \omega^{\tr(aj)}\ket{j},
\label{pauliopgal}
\end{eqnarray}
where $\omega$ is now defined as $\omega=e^{2\pi i/p}$, instead of $\omega=e^{2\pi i/q}$, and $\tr: GF(p^m) \rightarrow \mathbb{Z}_p$ is the trace function for a Galois field element given by
\begin{equation}
\tr(x) = x+x^p+x^{p^2}+...+x^{p^{m-1}} ,   \hspace{15 mm}   x \in GF(p^m)  .
\label{tracegalois}
\end{equation}
The above expressions may result in a different behavior for $X= X_1$ and $Z=Z_1$ operators as compared to the case of prime $q$. For example, $X$ and $Z$ will commute whenever $m$ is a multiple of $p$, as in $GF(4)$. More detail on above definitions can be found in Refs.~\cite{Ke06,Go14, Gr03}.

It can be shown that for the above generalized Pauli operators, the number of generators needed to build the stabilizer space of an $[\![n,0,d]\!]_{p^m}$ code is $mn$ \cite{Go14}. The key reason for this is that $GF(p^m)$ has the structure of a vector space of dimension $m$ over the field of $\mathbb{Z}_p$. 
The group of tensor products of $n$ $q$-dimensional Pauli operators, as defined in Eq.~\eqref{pauliopgal}, is then isomorphic to the corresponding Pauli group for $mn$ $p$-dimensional parties, where $p$ is prime. The latter is similar to the case considered in Sec.~\ref{coreconstruction}, for which $mn$ generators would be needed to describe the stabilizer space for $mn$ qudits of dimension $p$. It follows then $mn$ generators will also be needed to define $[\![n,0,d]\!]_{p^m}$.

Now, let us speculate a reduction-friendly form for the stabilizer representation in the case of power prime $q$. In Claim 1, for prime $q$ and $d \geq 2$, we showed that there were at least two generators in the set whose first Pauli element was not identity. We can extend this claim to the case of power-prime $q$ in the sense that now in every site there must be at least $2m$ generators with elements different from identity. This follows from distance property with respect to error operators of weight one, as well as the isomorphism we mentioned above, where $q$-dimensional qudits can be broken into $m$ $p$-dimensional qudits. For each of the latter qudits, we have already shown in Claim 1 that two non-identity operators are required on each site of their generator table. We can then argue that there would be at least $2m$ non-identity operators for $m$ $p$-dimensional qudits. 

Here is another observation that leads to the same conclusion. Let us focus on the first site. An error of weight one, on the first site, will take the form of $X_a Z_b$, with $a,b \in GF(p^m)$. Consider all generators of $[\![n,0,d]\!]_{p^m}$ with non-identity operators on their first site. Let us make a subgroup of Pauli operators by taking these generators to the power of all numbers in $\mathbb{Z}_p$ and multiply by each other. The resulting subgroup, let us call it $G_p$, will also contain elements in the form $X_a Z_b$, with $a,b \in GF(p^m)$. Let us define $A$ and $B$ to be, respectively, the set of numbers $a$ and $b$ for which $X_a Z_b \in G_p$. We can show that, unless $A$ and $B$ are equivalent to $GF(q)$, there is always an error operator with weight one that commutes with all generators. To show this, let us find dual bases $\{\alpha_i\}$ and $\{\beta_i\}$, $i=1,\ldots,m$, for $GF(q)$. That is, $\tr(\alpha_i \beta_j) = \delta_{ij}$, with $\delta_{ij}$ being the Kronecker delta function. Suppose the dimension of $A$ is strictly less than $m$. Then there exists an element $\alpha_k$ that does not belong to $A$. Operator $Z_{\beta_k}$ would then commute with all operators in $G_p$, and this violates $d \geq 2$. 

Based on the above observation, the non-identity operators on the first site must contain a basis for $\{X_a Z_b\}$, with $a,b \in GF(p^m)$. Such a basis set must contain $2m$ elements as shown above. A simple way of generating all elements in the form $X_a Z_b$, with $a,b \in GF(p^m)$, is to use a set of basis numbers $\{s_i\}$, $i=1,\ldots,m$, in $GF(q)$. $X_a$ can then be written as the product of powers in $\mathbb{Z}_p$ of $X_{s_i}$. The same holds for $Z_b$ and $Z_{s_i}$. For our reduction-friendly stabilizer form, we choose $s_i = \alpha^{i-1}$, $i=1,\ldots,m$. That is, we speculate that the generators of $[\![n,0,d]\!]_{q}$ can be written in the form given by Fig.~\ref{canstpower}. We now prove by construction that it is indeed possible to achieve this. 

\begin{figure}[h!]
\begin{tabular}{c | c c c c c | c}
& \multicolumn{5}{c}{$n/2$ sites} & $n/2$ sites \\
\hline
$g_1$ & $I$ & $I$ & $...$ & $I$ & $Z$ & $.................$\\
$g_2$ & $I$ & $I$ & $...$ & $I$ & $Z_\alpha$ & $.................$\\
$\vdots$ & $I$ & $I$ & $\vdots\vdots\vdots$ & $I$ & $\vdots$ & $\vdots\vdots\vdots\vdots\vdots\vdots\vdots\vdots\vdots\vdots\vdots\vdots\vdots\vdots\vdots\vdots\vdots$\\
$g_m$ & $I$ & $I$ & $...$ & $I$ & $Z_{\alpha^{m-1}}$ & $.................$\\
$g_{m+1}$ & $I$ & $I$ & $...$ & $I$ & $X$ & $.................$\\
$g_{m+2}$ & $I$ & $I$ & $...$ & $I$ & $X_\alpha$ & $.................$\\
$\vdots$ & $I$ & $I$ & $\vdots\vdots\vdots$ & $I$ & $\vdots$ & $\vdots\vdots\vdots\vdots\vdots\vdots\vdots\vdots\vdots\vdots\vdots\vdots\vdots\vdots\vdots\vdots\vdots$\\
$g_{2m}$ & $I$ & $I$ & $...$ & $I$ & $X_{\alpha^{m-1}}$ & $.................$\\
$\vdots$ & $\vdots$ & $\vdots$ & $\vdots\vdots\vdots$ & $\vdots$ & $\vdots$ & $\vdots\vdots\vdots\vdots\vdots\vdots\vdots\vdots\vdots\vdots\vdots\vdots\vdots\vdots\vdots\vdots\vdots$\\
$g_{n-2m+1}$ & $Z$ & $I$ & $...$ & $I$ & $I$ & $.................$\\
$g_{n-2m+2}$ & $Z_\alpha$ & $I$ & $...$ & $I$ & $I$ & $.................$\\
$\vdots$ & $\vdots$ & $I$ & $\vdots\vdots\vdots$ & $I$ & $I$ & $\vdots\vdots\vdots\vdots\vdots\vdots\vdots\vdots\vdots\vdots\vdots\vdots\vdots\vdots\vdots\vdots\vdots$\\
$g_{n-m}$ & $Z_{\alpha^{m-1}}$ & $I$ & $...$ & $I$ & $I$ & $.................$\\
$g_{n-m+1}$ & $X$ & $I$ & $...$ & $I$ & $I$ & $.................$\\
$g_{n-m+2}$ & $X_\alpha$ & $I$ & $...$ & $I$ & $I$ & $.................$\\
$\vdots$ & $\vdots$ & $I$ & $\vdots\vdots\vdots$ & $I$ & $I$ & $\vdots\vdots\vdots\vdots\vdots\vdots\vdots\vdots\vdots\vdots\vdots\vdots\vdots\vdots\vdots\vdots\vdots$\\
$g_{n}$ & $X_{\alpha^{m-1}}$ & $I$ & $...$ & $I$ & $I$ & $.................$\\
\hline
\end{tabular}
\caption{Reduction-friendly generator set for an $AME(n,q)$ for $n$ even and $q=p^m$ for $p$ prime: The left part, of length $n/2$, is set in this way and guarantees independence of the generators, whereas the remaining blank part, of length $n/2$, on the right must be found by imposing the remaining conditions of commutativity and distance. $\alpha$ is a primitive element of the Galois field $GF(p^m)$.}
\label{canstpower}
\end{figure}

{\it Proposition 3}. For an $AME(n,q)$ state, the generators of the stabilizer representation of the equivalent $[\![n,0,n/2+1]\!]_q$ code, with $n \geq 2$ even and $q=p^m$, with $p$ prime, can be brought to the following form

 \[ \begin{cases} 
      g_{i,j} = Z_{\alpha^k} & {\rm if} \hspace{2mm} i=2m(n/2-j)+(k+1) \\
      g_{i,j} = X_{\alpha^k} & {\rm if} \hspace{2mm} i=2m(n/2-j)+(m+k+1) \\
      g_{i,j} = I & {\rm elsewhere} 
   \end{cases}
\]

for $j = \{1,...,n/2\}$ and $k = \{0,1,...,m-1\}$, as shown in Fig.~\ref{canstpower}. \\

{\it Proof}. We proceed in parallel to the proof of Proposition 2, indicating the places where the treatment is different. Let us assume that the last $2m$ generators are the non-identity ones responsible for correcting weight-one errors on the first site. Consider the most general case, up to a constant factor, for those generators:
\begin{eqnarray} 
g_{n-r,1} = X_{a_r} Z_{b_r}, \hspace{5mm}  r=0,1,...,2m-1,
\end{eqnarray}
where $a_r, b_r \in GF(p^m)$. \\

{\it Step 1: Elimination of $Z_{b_r}$ to obtain $g_{n-r,1} = X_{a'_r}$}, for $r=0,\ldots,m-1$. As with the case of $q$ prime, we are free to substitute a generator by a power of itself multiplied by powers of the others, where all powers are in $\mathbb{Z}_p$. Let us make the following transformations sequentially starting with $r=0$, and then moving up to $r=m-1$. For $r=0$, we obtain
\begin{equation}
g'_{n} = \prod_{l=0}^{2m-1} (g_{n-l})^{\beta_{0l}},
\end{equation}
with $\beta_{0l} \in \mathbb{Z}_p$. In order to eliminate $Z_{b_0}$, we then require that
\begin{equation}
\sum_{l=0}^{2m-1} \beta_{0l} b_l = 0 \hspace{2mm} \rm{(mod \hspace{1mm} {\it p})}.
\end{equation}
Given that $b_l$'s are elements of $GF(q)$ of dimension $m$, there exists a non-trivial solution for $\beta_{0l}$ to satisfy the above equation, otherwise we would get a contradiction with the existence of $2m$ independent non-identity operators in the first place. Now that we removed $Z_{b_0}$, we can move to $r=1$, and repeat the same procedure using $g_{n-1}$ to $g_{n-2m+1}$. Continuing to higher values of $r$, at each step, we obtain $2m-r > m$ equations of the form 
%
\begin{equation}
\label{linind2}
\sum_{l=r}^{2m-1} \beta_{rl} b_l = 0 \mbox{ \ \ (mod {\it p}), \ \  $r=1,...,m-1$},
\end{equation}
with $2m-r > m$ variables, $\beta_{rl} \in \mathbb{Z}_p$, and known elements $b_l \in GF(p^m)$. Now, again given that $GF(p^m)$ is a vector space over $\mathbb{Z}_p$ with dimension $m$, Eq.~\eqref{linind2} will have non-trivial solutions, as it is effectively a linear combination of more than $m$ vectors, which have therefore to be linearly dependent. 

As in the prime case, if this first step produces $a_r'=0$ and therefore $g_{n-r}=I$, that means that we have to permute generators and select another set of $g_{n-r}$ (with $r=0,1,...,2m-1$) generators that are independent on the first site, which must exist for the stabilizer set to define a code of distance larger than 1.

{\it Step 2: Transformation of $X_{a'_r}$ to $X_{\alpha^{m-1-r}}$}. Here we state that as both the obtained $X_{a'_r}$ elements and the desired $X_{\alpha^{m-1-r}}$, for $r=0,1,...,m-1$, both form valid bases for the $\{X_a\}$ operators, with $a \in GF(q)$, there must exist a change of basis that takes us from one to another. Such a change of basis can be applied sequentially by starting from the bottom and then move up to the top. 

{\it Step 3: Transformation of the rest of the first column.} As with the prime case, the same procedure can be carried to transform $g_{n-r,1} = X_{a_r} Z_{b_r}$ to $g'_{n-r,1} = Z_{b'_r}$ and then to $g''_{n-r,1} = Z_{\alpha^{2m-1-r}}$, for $r = {m,m+1,...,2m-1}$. Then, we can easily make all the other generators to have identities on the first site by taking products of themselves with the last $2m$ generators, which form a basis of the Pauli group, each of appropriate power.

{\it Step 4: Iteration over the rest of the sites.} This step is also equivalent to the prime case, but now we have to eliminate each time the last $2m$ generators to obtain the child code. By iterating the procedure until reaching column $n/2$, we get the desired reduction-friendly form. $\square$ \\

As an example, Fig.~\ref{513422} shows the reduction-friendly form for $[\![5,1,3]\!]_9$ \cite{Go15b}. By removing the last four rows, we can then easily obtain the stabilizer form for the child code $[\![4,2,2]\!]_9$

\begin{figure}[h!]
\begin{tabular}{c | c c | c c c}
\multicolumn{6}{c}{$[\![5,1,3]\!]_9$}\\
\hline
$g_1$ & $I$ & $Z$ & $XZ_{-1}$ & $Z_{-1}$ & $X_{-1} Z$ \\
$g_2$ & $I$ & $Z_\alpha$ & $X_\alpha Z_{-\alpha}$ & $Z_{-\alpha}$ &$X_{-\alpha} Z_\alpha$\\
$g_3$ & $I$ & $X$ & $Z$ & $X_{-1}$ & $Z_{-1}$\\
$g_4$ & $I$ & $X_\alpha$ & $Z_\alpha$ & $X_{-\alpha}$ & $Z_{-\alpha}$\\
$g_5$ & $Z$ & $I$ & $Z_{-1}$ & $XZ_{-1}$ & $ZX_{-1}$\\
$g_6$ & $Z_\alpha$ & $I$ & $Z_{-\alpha}$ & $X_{\alpha}Z_{-\alpha}$ & $ZX_{-\alpha}$\\
$g_7$ & $X$ & $I$ & $X_{-1}$ & $Z$ & $Z_{-1}$\\
$g_8$ & $X_\alpha$ & $I$ & $X_{-\alpha}$ & $Z_\alpha$ & $Z_{-\alpha}$\\
\hline
\end{tabular}
\quad
\begin{tabular}{c}
$\rightarrow$
\end{tabular}
\quad
\begin{tabular}{c | c | c c c}
\multicolumn{5}{c}{$[\![4,2,2]\!]_9$}\\
\hline
$g_1$ & $Z$ & $XZ_{-1}$ & $Z_{-1}$ & $X_{-1} Z$ \\
$g_2$ & $Z_\alpha$ & $X_\alpha Z_\alpha$ & $Z_{-\alpha}$ & $X_{-\alpha} Z_\alpha$\\
$g_3$ & $X$ & $Z$ & $X_{-1}$ & $Z_{-1}$\\
$g_4$ & $X_\alpha$ & $Z_\alpha$ & $X_{-\alpha}$ & $Z_{-\alpha}$\\
\hline
\end{tabular}
\caption{Transformation from code $[\![5,1,3]\!]_9$ to its child code $[\![4,2,2]\!]_9$}
\label{513422}
\end{figure}

\section{Application to quantum repeaters}
\label{repeaters}

A one-way or third generation quantum repeater makes use of quantum error correction codes for correcting both loss and operational errors \cite{Mu15}. Quantum parity codes were studied first in this context \cite{Mu12,Mu14}, but later quantum polynomial codes (QPyC) \cite{Gl16,Mu17} and quantum Reed-Solomon codes (QRSC) \cite{Mu18} were shown to be more efficient. This improvement mainly comes from the fact that the latter two codes are QMDS codes, as shown in Fig.~\ref{venndiagram}. Being also CSS codes, there are known encoder/decoder circuits, which rely only on transversal operations on a maximum of two qudits at a time.

In this section, we investigate the importance of QMDS codes, especially those obtained from an AME state, in one-way repeater setups.  We do not consider the gate errors and only focus on the erasure errors, which result because of photon loss in the system and the channel. We calculate the number of equivalent qubits that can be transferred by a chain of $r$ repeater stations in which two adjacent nodes are apart by a distance $L_0$. In the first node, $k$ qudits of dimension $q$ are encoded to a codeword of length $n$ using a $[\![n,k,d]\!]_q$ code. Such a codeword is then transmitted by $n$ photons to the next node, at which point, the codeword is decoded back to $k$ qudits, with errors hopefully corrected, and then re-encoded again into an $n$-qudit codeword for transmission to the next node. The same procedure is repeated until we reach the final node. 

For such a repeater, the rate at which, in qubit per unit of time, quantum states are transferred via our repeater chain of $r$ links is given by
\begin{equation}
\label{Eq:rate}
R=k\frac{(P_{\rm success})^r}{t_0}\log_2q,
\end{equation}
where $t_0$ is the time taken for local operations at encoder-decoder modules, and \cite{Mu14,Mu15,Mu17,Mu18}
\begin{equation}
\label{Eq:PSucc}
P_{\rm success}=\sum_{j=0}^{d-1} {n\choose j} p_l^j (1 - p_l)^{n - j},
\end{equation}
where $p_l = 1-\eta_c e^{-\frac{L_0}{L_{att}}}$ is the probability of losing a physical qudit over a link of length $L_0$, and $L_{att}$ is the attenuation length of the channel, here assumed to be 20~km for current optical fibers. Parameter $\eta_c$ models the coupling efficiency in-and-out of encoder-decoder modules, which is assumed to be one in this section. Equation \eqref{Eq:PSucc} is the probability of successfully recovering the transmitted logical codeword encoded in the $[\![n,k,d]\!]_q$ code space, from one station to the next, taking only loss effects into account. We have assumed the codeword is transmitted using $n$ photons, each in one of $q$ possible modes. In Eq.~ \eqref{Eq:PSucc}, we use the fact that for a code of distance $d$, we can correct $d-1$ erasure errors.

To have a fairer comparison between different code structures used in a quantum repeater, it is common to define normalized cost factors that account for the amount of resources needed to get a particular rate, as, e.g., $R$ in Eq.~\eqref{Eq:rate}. Conventionally, this cost factor is defined as the number of qubits needed normalized by the rate $R$. Here, we are going to dig deeper into the issue of cost and define more practical cost parameters by distinguishing between short-term and long-term costs for a repeater system. We identify the short term cost as what is typically attributed to the initial investment cost, $C_{\rm initial}$, of a repeater setup i.e., the cost of the infrastructure that needs to be put in place to have a functioning repeater, whereas we attribute the long-term costs to the running cost of the system, $ C_{\rm running}$. The total cost over a time period $T$ could then be written as 
\begin{equation}
C_T = C_{\rm initial} + C_{\rm running}(T),
\end{equation}
during which $RT$ qubits have been transferred across the repeater chain.  Note that $C_{\rm initial}$ is not a function of $T$ as it represents the initial cost of setting up the repeater set up at a reference point.

In the absence of real estimates for the above cost functions, here we attempt to approximate their order of magnitude in terms of certain system parameters. We approximate $C_{\rm initial}$ by $N_{\rm rep} r \log_2(q) C_{\rm qubit}$, where $N_{\rm rep}$ is the number of qudits at each repeater station, which is expected to be proportional to $n$ and will depend on the concrete model of the repeater stations used, and $C_{\rm qubit}$ is the nominal cost for deploying the hardware corresponding to one qubit at the encoder/decoder modules. Similarly, we estimate the running cost, $ C_{\rm running}(T)$, by $n r q C_{\rm photon} T/t_0$, where $C_{\rm photon}$ is the cost associated with processing and transmission of a single photon over an elementary link of length $L_0$. Here, we have made two assumptions. First, that we have used the system at its maximum possible rate by sending one codeword through the system every $t_0$ unit of time. Secondly, we have assumed that the cost is proportional to $q$, as, in our model, we need that many modes (channels) to transmit a single photon.

Using the above estimates, the total cost over a running time of $T$, per $RT$ transmitted qubit, per unit of distance is given by
\begin{equation}
\frac{C_T}{RT rL_0} \approx \frac{N_{\rm rep}  \log_2(q)}{L_0 R} C_{\rm qubit}/T + \frac{n q}{L_0 R} C_{\rm photon} /t_0.
\end{equation}
Assuming that $C_{\rm qubit} \gg C_{\rm photon}$, for small values of $T$, the first term in the above equation is the major source of cost, but as $T$ becomes larger and larger, one may ignore the first term, as the second term would contribute most to the total cost. Based on this observation, here we introduce two cost coefficients: 
\begin{equation}
C_{\rm ST} = \min_{L_0} \frac{n \log_2(q)} {L_0 R t_0}, 
\label{eqcosts}
\end{equation}
which represents the short-term investment cost factor, whereas 
\begin{equation}
C_{\rm LT} = \min_{L_0} \frac{n q} {L_0 R t_0}, 
\label{eqcostl}
\end{equation}
which represents the long-term running cost factor. In \eqref{eqcosts}, we have assumed $N_{\rm rep} \propto n$, and have normalized the coefficient by the constant time $t_0$ to have the same dimension as $C_{\rm LT}$. Also note that $C_{\rm ST}$ is independent of $q$, as $R \propto \log_2(q)$. We use the above cost coefficients to compare different codes used in quantum repeater setups taking into account both the generated rate and the amount of resources needed. Note that if we are comparing codes of the same dimension $q$, it is indifferent which cost factor is used for comparison. 
%
%
%
\begin{figure}[h!]
\centering
\includegraphics[width = 0.9\linewidth]{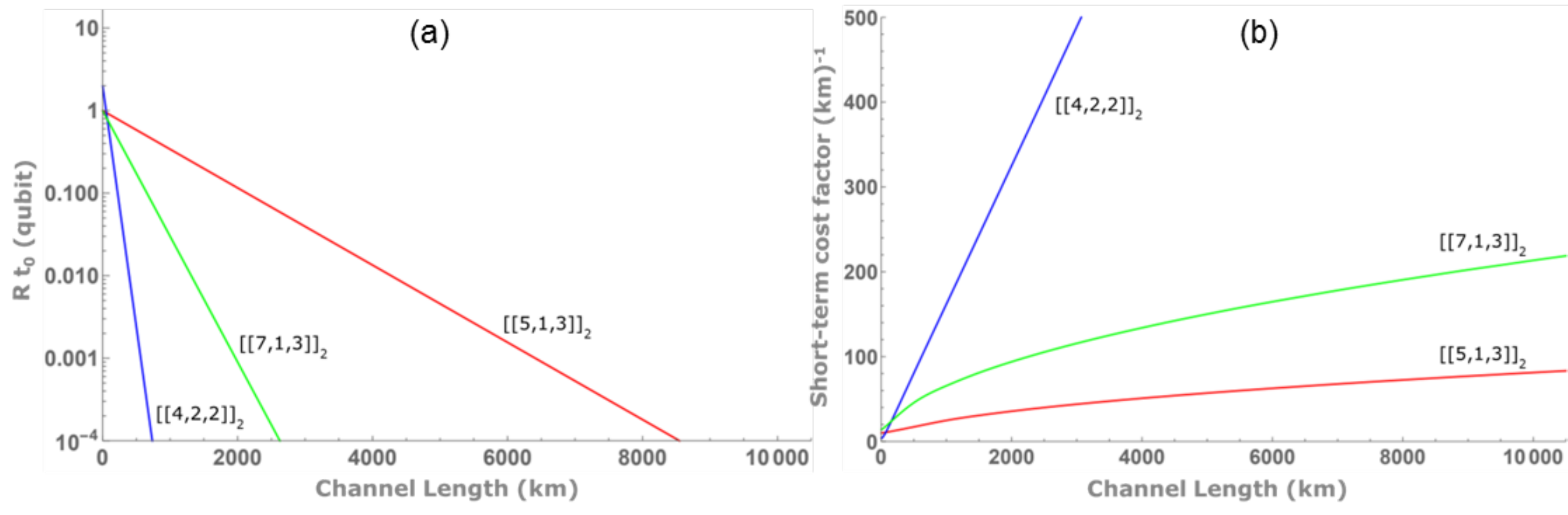}
\caption{(a) Average number of qubits transferred over a time period $t_0$, $Rt_0$, at $L_0 = 1$~km, and (b) short term cost coefficient, $C_{\rm ST}$, at the optimal value of $L_0$, versus total channel length for the children codes of AME(6,2) in comparison to the non-QMDS Steane code $[[7,1,3]]_2$ .   }
\label{Fig:AME62}
\end{figure}

\begin{figure}[h!]
\centering
\includegraphics[width = 0.9\linewidth]{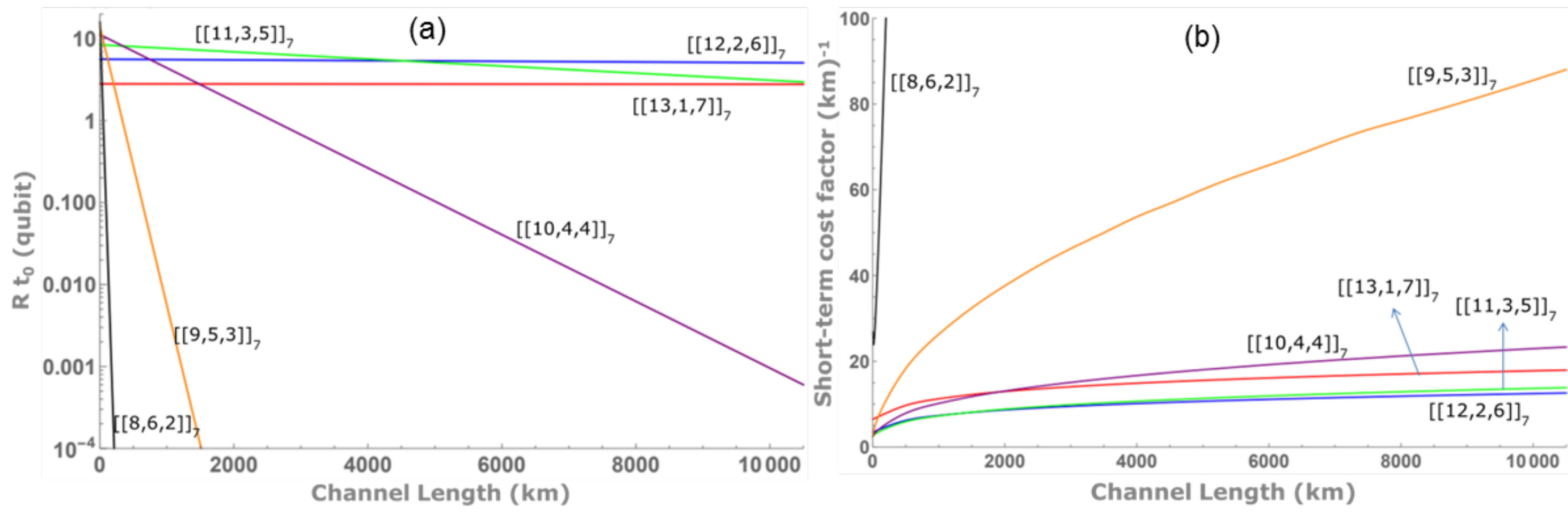}
\caption{(a) Average number of qubits transferred over a time period $t_0$, $Rt_0$, at $L_0 = 1$~km, and (b) short term cost coefficient, $C_{\rm ST}$, at the optimal value of $L_0$, versus total channel length for the children codes of AME(14,7). }
\label{Fig:AME147}
\end{figure}

Based on the above figures of merit, we have compared several QMDS families that emerge from the children codes of AME states. In Fig.~\ref{Fig:AME62}, we have considered the AME(6,2) state and have compared its two children codes in terms of the achievable rate (Fig.~\ref{Fig:AME62}(a)) and the short-term cost factor (Fig.~\ref{Fig:AME62}(b)). In our comparison, we have also included the non-QMDS Steane code. It is clear that, except for very short distances, the QMDS code $[\![5,1,3]\!]_2$ offers the best performance. This is the code with maximum distance possible at $n=5$, hence the minimum value of $k=1$. It is interesting that this is not necessarily the case for all values of $n$. In fact, as shown in Fig.~\ref{Fig:AME147}, for children codes of AME(14,7), $[\![12,2,6]\!]_7$ and $[\![11,3,5]\!]_7$ offer the lowest cost factors at all distances considered. They also beat the sibling code with maximum distance, $[\![13,1,7]\!]_7$, in terms of achievable rate. This behavior can be attributed to the fact that, when $n$ is sufficiently large, the optimal code just needs to have a sufficiently large $d$ to combat erasure errors. Once $d$ reaches that level, we can maximize $k$, while satisfying QMDS condition, to improve the achievable rate in our system. It is also noteworthy that, at very short distances, the optimal code could be different from that obtained at long distances. This could be an important observation for applications in distributed quantum computing, where, initially, the objective is to connect several few-qubit quantum processors at short distances from each other.


One can generalize the above comparison and carry it out for all known AME states. In Table~\ref{compame}, we have done this for all AME states with $n\leq 14$ and $q\leq 8$, taken from Refs.~\cite{Hu18a,Hu18b}, according to their long-term cost coefficients. For each AME state, we have first found, among all children codes of that AME state, the code that gives the minimum long-term cost at a total channel length of 1000~km and 10,000~km. Note that the same code would also offer a minimum short-term cost factor as parameters $n$ and $q$ are the same for all children codes. We have specified the value of $k$ that corresponds to the optimal code for each cell in the table. In each cell, the number on the left corresponds to a channel length of 1000~km, and the number on the right is for the 10,000~km case. The optimal value of $k$ for each row $n$ and column $q$ would then specify the code $[[n-k,k,\floor{n/2}+1-k]]_q$ as the best within its family. We have also identified the optimal code for a fixed $q$, as well as for a fixed $n$. These codes have been highlighted, respectively, by light blue and light red colors. If the optimal code in a row and a column coincide, we have highlighted it with purple.


\begin{table}
\begin{tabular}{c | c | c | c | c | c | c | c }
\backslashbox{$n$}{$q$} & 2 & 3 & 4 & 5 & 6 & 7 & 8 \\
\hline
4 & - & \cellcolor{LightCyan} 1,1 & 1,1 & 1,1 & ? & 1,1 & 1,1 \\
\hline
5 & 1,1 & \cellcolor{LightCyan} 1,1 & 1,1 & 1,1 & 1,1 & 1,1 & 1,1 \\
\hline
6 & \cellcolor{LightRed} 1,1 & \cellcolor{LightCyan} 1,1 & 1,1 & 1,1 & \cellcolor{LightRed} 1,1 & 1,1 & 1,1 \\
\hline
7 & - & \cellcolor{LightCyan} 1,1 & ? & 1,1 & ? & 1,1 & 1,1 \\
\hline
8 & - & - & ? & \cellcolor{LightCyan} 1,1 & ? & 1,1 & 1,1 \\
\hline
9 & - & \cellcolor{LightCyan} 1,1 & 1,1 & 1,1 & ? & 1,1 & 1,1 \\
\hline
10 & - & \cellcolor{LightPurple} 2,1 & \cellcolor{LightRed} 2,1 & \cellcolor{LightRed} 2,1 & ? & 2,1 & 2,1 \\
\hline
11 & - & ? & ? & ? & ? & \cellcolor{LightCyan} 2,1 & ? \\
\hline
12 & - & - & - & ? & ? & \cellcolor{LightCyan} 3,2 & 3,2 \\
\hline
13 & - & - & ? & ? & ? & \cellcolor{LightCyan} 3,2 & ? \\
\hline
14 & - & - & ? & ? & ? & \cellcolor{LightPurple} 3,2 & \cellcolor{LightRed} 3,2 \\
\hline
\end{tabular}
\caption{The optimal value of $k$ that minimizes $C_{\rm LT}$ for children codes originated from $AME(n,q)$. For each $n$ and $q$, the number on the left of the respective table cell corresponds to the optimal $k$ for a total channel length of 1,000~km, whereas the one on the right corresponds to 10,000~km. The optimal code is then given by $[[n-k,k,\floor{n/2}+1-k]]_q$. Cyan coloured codes indicate the best ones of each row, light red coloured codes indicate the best of the column, and purple coloured ones indicate the best in both their column and row. A dash (-) indicates that an AME state does not exist for those parameters, and a question mark (?) indicates that it is not known whether an AME state exists. }
\label{compame}
\end{table}

There are several observations that can be made from Table~\ref{compame}. First, note that, as pointed out before, for small code lengths $n$, the optimal code only encodes one qudit, i.e., $k=1$. This starts to change when $n\geq 10$, especially, first, at shorter distances. The other interesting point is that the optimal code in each column is the one with largest $n$, which suggests that the overall cost would be lower if we have the technology to reliably generate larger codewords. Interestingly, with the exception of the rows corresponding to $n=5,6$, which are the only ones for which a family of QMDS qubit codes exist, the best code for a fixed $n$ happens at lowest value of $q$ possible. This is important as practically it is harder to work with high-dimensional systems. The function $q/\log_2(q)$, for an integer $q$, takes its minimum value at $q=3$, and is monotonically ascending for $q>3$. This is why $q=3$, or the minimum possible $q$, minimize the long-term cost factor for $n>6$.

Before finishing this section, let us revisit the results reported in Ref.~\cite{Mu18}, in which the short-term cost coefficients for QPC, QPyC and QRSC codes are compared in the absence of operational errors. There, the authors conclude that QRSC are the best ones by a large factor, without specifying which codes in particular have been used in their analysis. As QPyC codes have $k=1$ and QRSC codes have $k\geq 1$, and both of them are QMDS, we believe the above conclusion would hold only if sufficiently large codewords are in use. Their finding would then be in agreement with what we find in our analysis with respect to the code size $n$, and the encoded number of qudits, $k$.

Finally, note that we have kept here the analysis simple by not considering operational errors. If we want to include them, we should consider a concrete repeater model to estimate the global error $\epsilon$ out of the individual sources of error. For TEC stations and CSS codes, $\epsilon \simeq (3\epsilon_g + 4\epsilon_d + \epsilon_m)$ \cite{Mu18}, where $\epsilon_g$ stands for gate errors, $\epsilon_d$ for depolarization errors and $\epsilon_m$ for measurement errors. For non-CSS codes with a different implementation, $\epsilon$ could be significantly different. Error analysis for non-CSS codes is beyond the scope of this paper, however, and shall be addressed separately.

\section{Conclusions}
\label{Sec:Diss}

In this work, we provided construction techniques by which we could start with the stabilizer representation of an AME state, and obtain the stabilizer form for all its QMDS children codes. This was achieved by introducing a reduction-friendly form for the stabilizer generators in the case of systems with prime or power-prime dimensions. Such a reduction-friendly form could, in principle, help with finding new AME states and QMDS codes as half of the columns of the generators are specified by our construction technique. We then used the new AME-based classification of the resulting QMDS codes and studied the applicability of such codes in one-way quantum repeaters. We specifically considered short-term and long-term cost factors that accounted for both the performance and the required resources of quantum repeaters. We came up with general guidelines on how to choose the optimal code for repeater applications. In particular, we found that, so long as the major source of error is the loss-driven erasure errors, we should choose codes with largest possible size and lowest possible dimension (with minimum three). The optimal number of qudits to encode, $k$, would then be specified by the length of the code and the distance at which we need to communicate. The larger the former and the shorter the latter, the higher the optimal value of $k$ would be. We note that, in our analysis, we did not consider the effect of processing errors. This is particularly subtle because the more efficient QMDS codes may not have a CSS form, in which case their encoding and decoding would be non-trivial, especially, in higher dimensions. This requires us to develop new techniques to perform complete error analysis on such systems.

\section*{Acknowledgements}
We thank F. Fit\'e for the insights given in Galois fields, F. Huber for careful reading and new proposals, and Y. Jing, O. G\"uhne, D. Goyeneche, A. Riera, M. Grassl, S. Muralidharan, C. Gogolin, Z. Raissi, G. Curr\'as and J. I. Latorre for interesting discussions and corrections. This work is partially funded by the UK EPSRC Grant No. EP/M013472/1. All data generated in this paper can be
reproduced by the provided methodology and equations.

\bibliographystyle{apsrev4-1}
\bibliography{danialsina}

\end{document}